\documentclass[structabstract]{aa}
\usepackage{graphicx}
\usepackage{natbib}
\usepackage{multirow}
\usepackage{tabularx}
\usepackage{amssymb}

\bibpunct{(}{)}{;}{a}{}{,} 

\begin{document}

   \title{Multiperiodicity, modulations and flip-flops in variable star light curves\thanks{Based on 
observations made with 
the Nordic 
Optical Telescope, operated
on the island of La Palma jointly by Denmark, Finland, Iceland,
Norway, and Sweden, in the Spanish Observatorio del Roque de los
Muchachos of the Instituto de Astrofisica de Canarias. }}
   \subtitle{III. Carrier fit analysis of LQ Hya photometry for 1982-2014}
   \authorrunning{N. Olspert et al.}
   \titlerunning{LQ Hya photometry}

   \author{N. Olspert \inst{1} \and 
           M. J. K\"apyl\"a \inst{1} \and
           J. Pelt \inst{2} \and
           E. M. Cole \inst{3} \and
           T. Hackman \inst{3,4} \and
           J. Lehtinen \inst{3} \and
           G. W. Henry \inst{5}}

   \offprints{N. Olspert\\
          \email{nigul.olspert@aalto.fi}
          }   

\institute{
ReSoLVE Centre of Excellence, Aalto University, Department of Information and Computer Science, PO Box 15400, FI-00076 Aalto, Finland
\and Tartu Observatory, 61602 T\~{o}ravere, Estonia \and
Department of Physics, PO Box 64, FI-00014 University of Helsinki, Finland \and
Finnish Centre for Astronomy with ESO (FINCA), University of Turku, V\"{a}is\"{a}l\"{a}ntie 20, FI-21500 Piikki\"{o}, Finland \and
Center of Excellence in Information Systems, Tennessee State University, 3500 John A. Merritt Blvd., Box 9501, Nashville, TN 37209, USA
}

\date{Received / Accepted}

\abstract{}
{We study LQ Hya photometry for 1982-2014 with the carrier fit (hereafter CF) -method
and compare our results to earlier photometric analysis and recent Doppler imaging 
maps.}
{As the rotation period of the object is not known a priori, we first utilize different types of statistical methods (least-squares fit of harmonics, 
phase dispersion statistic) to estimate various candidates for the carrier period for the CF method. 
Secondly, a global fit to the whole data set and local fits to shorter segments are computed with the period that is found to be the optimal one.
}
{
The harmonic least-squares analysis of all the available data reveals a short period close to 1.6 days as a limiting value for a set of
significant frequencies. We interpret this as the rotation period of the spots near the equatorial region. 
In addition, the distribution of the significant periods is found to be bimodal, hinting of a longer-term modulating period, which 
we set out to study with a two-harmonic CF model. Weak modulation signal is, indeed retrieved, with a period of roughly 6.9 years.
The phase dispersion analysis gives a clear symmetric minimum for coherence times lower than and around 100 days. We interpret this as the mean rotation pattern of the spots.
Of these periods, statistically the most significant and physically most plausible is the mean spot rotation period 1{\fd}60514, which is chosen
to be used as the carrier period for the CF analysis.
With the CF method we seek for any systematic trends in the spot distribution in the global time frame, and locally look for abrupt phase changes earlier reported in
rapidly rotating objects.
During 2005--2008 the global CF reveals a coherent structure rotating with a period of 1{\fd}6037, 
while during most other times the spot distribution appears rather random in phase. 
}
{The evolution of the spot distribution of the object is found to be 
very chaotic, with no clear signs of an azimuthal dynamo wave that
would persist over longer time scales, although the short-lived
coherent structures observed occasionally do not rotate 
with the same speed as the mean spot distribution. 
The most likely explanation of the bimodal period distribution is attributed to the high- and low latitude spot formation regions
confirmed from DI and ZDI.
}

 
\maketitle

\section{Introduction}
 
\object{LQ Hya} (\object{HD 82558}, \object{GL355}) is a 
chromospherically active BY Draconis
-type star of the spectral type K2V
\citep{cutispoto1991,covino2001}. It also shows a high level of Ca H\&K
emission \citep[log\,$R'_{\rm HK}$=-4.06,][]{white2007}, manifesting
very high level of magnetic activity.  With an estimated mass of
0.8$\pm$0.1\,M$_{\odot}$ and age 51.0$\pm$17.5 Myrs
\citep{Tetzlaff2011}, the star is considered a young solar
analogue. The star spins very fast, with the estimated rotation period
being around 1.6 days
\citep[e.g.][]{jetsu1993,berdyugina2002,kovari2004,lehtinen2012}.

In addition to exhibiting strong magnetic activity indicators, the star
shows modulation in its light curve, as first proposed by
\citet{eggen1984} and confirmed by \citet{fekel1986}.  Such behavior
is interpreted as cool spots rotating with the stellar
surface. 
For this reason
photometric light curves have been used to determine the rotation
period of the star.  If the star exhibits latitudinal and/or radial
surface differential rotation analogous to the Sun, or latitudinal
dynamo waves (which is the solar case, as the sunspots form the
well-known butterfly diagram with cyclic behavior known as 
Sp\"orer's law) and/or azimuthal dynamo waves \citep[which can occur
in the rapid rotation regime, e.g.][]{lindborg2011}, this simple
picture might not be applicable.
Indeed, based on previous studies of LQ Hya photometry it has become evident 
that no single period suitable for describing the physical system throughout 
the whole observational time span available exists. 
However, for shorter epochs dominating periods have been found.
In \citet{jetsu1993} a good phase coherence was achieved with a period of $1{\fd}601136$,
\citet{berdyugina2002} arrived at a period estimate of $1{\fd}601052$, \citet{kovari2004}
report a period of $1{\fd}60066$, and the analysis of \citet{lehtinen2012} gives a period $1{\fd}6043$.

The surface differential rotation of the object has been estimated
by either using photometric lightcurves
\citep{jetsu1993,berdyugina2002,you2007,lehtinen2012}, 
or spectroscopic observations
analyzed with Doppler imaging (hereafter DI) methods
\citep{strassmeier1993,saar1994,rice1998,kovari2004} 
and Zeeman Doppler imaging (hereafter ZDI)
methods \citep{saar1994,donati1999,donati2003,mcivor2004}. 
It is customary to define the
differential rotation parameter as
\begin{equation}
k=\frac{\Omega_{\rm eq}-\Omega_{\rm pole}}{\Omega_{\rm eq}}=1-\frac{\Omega_{\rm pole}}{\Omega_{\rm eq}},
\end{equation}
describing both the magnitude and the type of the latitudinal rotation
law. Large values of $k$ denote strong differential rotation, positive
signs corresponding to solar-like profiles with a faster equator and a
slower pole, negative to anti-solar profiles with faster poles and a
slower equator. From photometry, only the magnitude of $k$ can be
deduced, while using DI one
can also determine the sign of it. 
The values obtained from the fluctuations in the photometric period range from 
$k=0.015$ \citep{jetsu1993} to 0.025 by \citep{you2007}.

The DI and ZDI results of \citet{saar1994} initially estimated an upper 
limit of  differential rotation based on polar `smearing' of 
$k \lesssim 0.03$ and further analysis
by \cite{kovari2004,donati2003b} indicate
even weaker ($k=0.002...0.006$) solar-like differential rotation. Due to
the relatively low $v \sin(i)$ of the object versus the required value for DI
techniques, the results show significant scatter \citep[see
e.g.][]{barnes2005}. Obtaining a reliable value for differential
rotation using DI and ZDI requires a longer baseline of observations
than used in most studies \citep{strassmeier1993,rice1998}.

In general, the amount
of differential rotation in this object can be concluded to be very
small compared to the solar value of $k \approx 0.2$. 
Hydrodynamical mean-field modeling of the rotation law of this
object by \citet{kitchatinov2011} agrees with the observations in the
sense that the obtained profiles are solar-like, and the magnitude
falls within the observational range ($k=0.028$).

The DI and ZDI maps
\citep[e.g.][]{strassmeier1993,saar1994,rice1998,donati1999,
donati2003,cole2014b} show both high- and low-latitude spot activity
on the object to the extent that \citet{cole2014b} define the
distribution over latitude to be bimodal. The relative strength of
the two latitudinal spot regions has been reported to be highly
variable over time so that during some epochs the near-equator spots
dominate while during others the nearly polar features are the
strongest. The spot distribution from DI and ZDI has been postulated
to be concentrated onto active longitudes during some epochs, therefore
being occasionally highly non-axisymmetric \citep[e.g.][]{saar1994},
but during different epochs no clear signs of active longitudes have
been found \citep[e.g.][]{cole2014b}.

Many authors of photometric or DI studies have reported the
concentrations of the spot activity on certain longitudes 
\citep{jetsu1993,berdyugina2002,kovari2004,lehtinen2012}. The periods producing the
largest amount of phase clustering and the phase separation of the `active longitudes' vary 
significantly
depending on the data set span and timing, and also the method used, indicating that these structures
are not persistent, but rather may be changing strongly in time. For example, \citet{lehtinen2012}
concluded that during 1988-2012, two different periods describing structures with active longitudes
$P_{\rm al}=1{\fd}61208$ and $P_{\rm al}=1{\fd}603693$, the former
one appearing around 1995 and the latter one between the years 2003 and
2009, were needed to describe the phase clustering of the data.
In contrast, 
\cite{berdyugina2002} used a light curve inversion technique to recover 
spot phases and postulated the existence of two active longitudes 
about $180\degr$ apart, spanning the 
entire 20 years of the data set, and calculated a new rotation period 
for the spot structure from 
the drift of these active longitudes, $P_{\rm al}=1{\fd}601052 
\pm 0{\fd}000014$.

The mean brightness of the star exhibits obvious signs of cyclic activity.
Various determinations from photometry indicate cycles of around 6-7 years
\citep[e.g.][]{jetsu1993,strassmeier1997,cutispoto1998,olah2009}, while also
co-existing shorter \citep[of the order of 3 years, e.g.][]{messina2003} and longer 
\citep[of the order of 11 years, e.g.][]{olah2000} cycles have been reported. 
\citet{berdyugina2002}, who postulated coherent active longitudes
during the time span of almost 20 years, found a 7.7-year cycle in the mean brightness,
and additionally a 5.2-year cycle related to the regular change of the activity
level of the two active longitudes (flip-flop). 

From the various earlier studies it is evident that the behavior of the object
is extremely complex, and a systematic approach to understand these complexities
is yet missing in the literature. This is attempted in the current study using
the 
CF method, the essential properties of which are
explained in Sect.~\ref{CFMO}. This method has been previously successfully 
applied to study spot activity
in other types of stars: see \citet{hackman2013} for the analysis of FK Coma
Berenices and \citet{lindborg2013} for the analysis of II Pegasi.
As the rotation period of the star is not known {\it a priori}, we first
search for the optimal carrier frequency using different kinds of statistical methods
described in Sect.~\ref{LSF}-\ref{D2}.
We then perform a global CF using the
optimal carrier frequency (Sect.~\ref{resGRref}),
the aim being to search for any persistent trends and/or phase disruptions.
Finally, we perform local CF-s to study local segments
in an attempt to identify interesting phase behavior. Here we also make a
comparison to the Continuous Period Search (hereafter CPS) method results by \citet{lehtinen2012} 
and the recent DI results of \citet{cole2014b}. In Sect.~\ref{disc} and 
Sect.~\ref{conc} we discuss and conclude our findings.

\section{Data}\label{data}

\begin{figure}
\begin{center}
\includegraphics[width=\columnwidth]{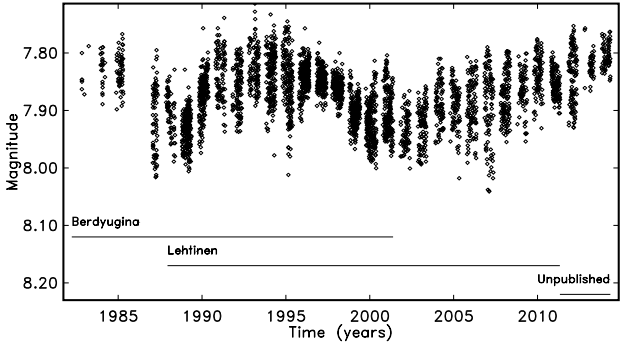}
\caption{The combined data set from the three different sources.
The dataset referred to as D1 consists of all available data, while dataset D2 comprises only T3-APT data (Lehtinen + unpublished).
}\label{datasets}
\end{center}
\end{figure}

In this paper we use the data consisting of nearly 32 years
of photometry from three different sources, namely the photometry
collected and published in \citet{berdyugina2002} from HJD = 2 445 275
(2 November 1982) to HJD = 2 452 053 (23 May 2001), the published
photometry of \citet{lehtinen2012} from HJD = 2 447 141 (11 December
1987) to HJD = 2 455 684 (2 May 2011) obtained with the T3 0.4m APT
at the Fairborn Observatory, Arizona and finally unpublished
photometry obtained with the same telescope from HJD = 2 455 685 (3
May 2011) to HJD = 2 456 783 (5 May 2014). From these data sources (see Fig.~\ref{datasets}) we
compiled two input data sets. First we rescaled T3-APT data to fit the
data from \citet{berdyugina2002} and 
for overlapping observations, computed data points as averages
(maximum allowed time difference of 0.1 days was used).
As a result of this procedure we got a long data set (hence D1)
consisting of 3929 observations covering 11508 days. This data set was
used to perform periodicity analysis and optimal carrier period value
estimation, the analysis presented in Sect.~\ref{carrier}. 
For the CF analysis presented in Sect.~\ref{res}, we combined all data from T3-APT
telescope, i.e. the data published in \citet{lehtinen2012} supplemented with the
seasons up to May 2014. As a result we got a homogeneous data set
(hence D2) which consisted of 2907 observations covering 9642
days. We denote this data set as homogeneous, as it was observed with
the same instrument with the same comparison star (HD 82428). We also
note that it is of better quality than the D1 data.


\section{Carrier fit method}\label{CFM}
\subsection{Overview}\label{CFMO}
A detailed description and the discussion of applicability of the CF method can be found in \citet{pelt2011}; 
here we briefly cover some of the key aspects of it. A CF model can be described as a truncated slowly modulated harmonic
decomposition of the signal:
\begin{equation} \label{CFmodel}
f(t) = a_{0}(t) + \sum\limits_{k = 1}^{K} (a_{k} (t) \cos (2 \pi t k\nu_{0} ) + b_{k}(t) \sin (2 \pi t k \nu_{0} )),
\end{equation}
where $\nu_0$ is the carrier frequency, $a_0(t)$ is the time-dependent
mean level of the signal, $K$ is the total number of harmonics
included in the model, describing the overtones of the basic carrier
frequency. $a_k(t)$ and $b_k(t)$ are the low-frequency signal
components which can be modeled by either splines or harmonics. 
Depending on the time series in question, either one of these approaches 
may be a more suitable choice, e.g. 
the spline approximation might be more suitable for cases where the signal is 
known to abruptly change. 
However, generally the difference between goodness of fits (for definition, see Sect.~\ref{CFSFP}) 
is marginal 
in case the number of free parameters for both methods are of the same order.
In the scope of the current study we limit ourselves to the option of harmonic modulators.

The value for the carrier frequency used in the above model can be, for instance,
the rotation period of the object, or any other clocking frequency describing the system.
In case of LQ Hya, however, the rotation period is not directly known, as it is
a single star. 
Therefore, we discuss various approaches to carrier frequency (or period) selection below.  

Following the notation used in \citet{pelt2011}, the trigonometric 
approximation of the 
slow amplitude modulation curves can be written as:
\begin{equation}
a(t) = c_0^a  + \sum\limits_{l = 1}^L {\big ( c_l^a\cos (2\pi tl\nu_D ) + s_l^a\sin (2\pi tl\nu_D )\big ),}
\end{equation}
\begin{equation}
b(t) = c_0^b  + \sum\limits_{l = 1}^L {\big ( c_l^b\cos (2\pi tl\nu_D ) + s_l^b\sin (2\pi tl\nu_D )\big ),} 
\end{equation}
where $L$ is the total number of harmonics used in the modulator
model and $\nu_{D}= 1/D=1/C\cdot(t_{\max}-t_{\min})$, where $D$ is data period, $C$ is the coverage factor 
and $[t_{\min},t_{\max}]$ is the time
interval to be fitted with the model. Here we note that the data period $D$
must be significantly longer than the carrier period $P_0=1/\nu_0$, 
and preferably also be a little 
longer than the actual 
time span of the data, i.e. the coverage factor should be $C \gtrapprox 1$.

\subsection{Selection of free parameters}\label{CFSFP}
One of the crucial things to consider when applying the CF method to real
data is the suitable values of the free parameters ($\nu_0$, $K$, $L$
and $C$).
As the selection of the carrier frequency $\nu_0$ is particularly important we dedicate the whole of the Sect.~\ref{carrier} to it.
The methods used for determining the values for other parameters will be discussed here.
In general we need to run the computations with all
possible combinations of parameter values drawn from some meaningful
ranges and then estimate the goodness of fit for every run using
\begin{equation} \label{rsquared}
R^2=1-\frac{\sum\limits_{i=1}^n\left(y_i-f_i\right)^2}{\sum\limits_{i=1}^n\left(y_i-\overline{y}\right)^2},
\end{equation}
where $n$ is the number of data points, $y_i$ is the value of the $i$-th data point, $f_i$ is the value
of the fit corresponding to the time moment of $i$-th data point and
$\overline{y}$ is the mean of the values of all data points. 
Qualitatively speaking the goodness of fit compares the variance of data points
around the model to that one of the data itself.
In the current study we use two approaches: we start by analyzing the full
set of data as a whole (global fit) followed by the
analysis of seasonal data segments (local fit). In both
cases we aim at as high $R^2$ values as possible while avoiding the
possibility of either overfitting the data or fitting into the gaps.

Before continuing with the methods of parameter selection we note that
the suitable value for $L$ is first of all dependent on the 
value of the coverage factor $C$, which 
defines the period of the slowest modulator in the model. 
It is reasonable to adopt a value of the same 
order or little longer than the length of the whole data set. 
This way we guarantee that the slowest detectable changes in the data are taken into account by the model.
In the current study we have fixed $C=1.2$.

The difficulty introduced by the gaps in the data constitutes the so called cycle count problem. 
The low frequency modulators $L$ introduce variance around the carrier frequency.
Here, we need to keep in mind that the difference in cycle counts for these maximum and minimum 
frequencies during the longest gap in the data should be less than one to 
avoid phase match indeterminacy.
This can be concisely expressed by the following criterion: $(\nu_0 + \nu_D)\Delta_{\rm gap} - (\nu_0 - \nu_D)\Delta_{\rm gap} < 1$, 
where $\nu_0$ is a high frequency carrier, $\nu_D$ is a low frequency modulator and $\Delta_{\rm gap}$ is the length of the longest gap in the given data set. 
After making replacement $\nu_D=L/D=L/C\cdot(t_{\max}-t_{\min})$ and simplifications we obtain an estimate for the upper limit of $L$:
\begin{equation} \label{cyclecount}
\frac{2L \cdot \Delta_{\rm gap}}{C\cdot(t_{\max}-t_{\min})} < 1 \Rightarrow L < \frac{C\cdot(t_{\max}-t_{\min})}{2\cdot\Delta_{\rm gap}}.
\end{equation}
Using this formula we can estimate the maximum valid $L$ for the data set with the given time span and the longest gap.

In case of local fits, due to the low number of data points the possibility of an
overfit appears. We use Bayesian information criterion (hereafter BIC) to determine the optimal values for $K$ and $L$ 
in a similar way it was done in \citet{lehtinen2011}, the differences being that we have 2 parameters in the 
model and we omit the weights of the data points. More precisely, we search for the minimum of the following criterion:
\begin{equation} \label{bic}
R_{BIC} = n \ln(\sigma^2) + ((2 + 4L) K + 2L + 1) \ln n,
\end{equation}
where $\sigma^2=\frac{1}{n-1}\sum\limits_{i=1}^n\left(y_i-f_i\right)^2$ and we have used the same notation as in Eq.~(\ref{rsquared}).
The first term of this equation describes the quality of the fit and the second one adds a penalty 
proportional to the total number of parameters in the model. 
Now, as we need the information for possible primary and secondary minima to 
be able to detect flip-flop type events, we omit $K=1$ from the set of trial values. 
For $L$ we don't impose a lower limit, but we will calculate the upper limit using Eq.~(\ref{cyclecount}) to check if the
value obtained from BIC is valid. If it exceeds the upper limit, we will use the latter one as the optimal value.
The practical application of this procedure is detailed in Sect.~\ref{resLF}.

In case of the global fit the possibility of overfitting is quite low
as the presence of long seasonal gaps in the data significantly lowers the maximum possible
values for $K$ and $L$. It is quite probable that when increasing the number of parameters,
the model starts showing big distortions in the regions where data is
missing, considerably earlier than high value of $R^2$ (e.g 90 \%) is achieved.

Before starting to measure the effect of the gaps on the
model, we first need to specify how long a region without data
qualifies as a gap. In our case the data is divided into observational
seasons where relatively densely spaced data is alternating with a bit
shorter ranges with no data at all. Based on a closer look at the
actual spacing of the data we define the gap as any region without
data that is longer than 130 days. 
This definition leads to 27 segments with data and 26 gaps
between them (minimum being 139 and maximum 302 days). The length of the
homogeneous data set is 9642 days, thus using Eq.~(\ref{cyclecount}) with the maximal gap size of 302 days we obtain
a maximum value for $L$ to be 19. As explained in Sect.~\ref{LSF} this is too low value to cover the full spectrum
of the data. Here we stand in front of the question either to leave out the data points preceding the longest gap
or lose the reliability of the model during this single gap while still obtaining better overall results. 
In the current study we decided to choose the latter option. The second longest gap in the data is 192 days long, 
increasing the suitable value for $L$ to 30. This number is already in the same order as the
number of observing seasons, so that we could expect good approximation of the seasonal variation 
by the term $a_{0}(t)$ in Eq.~(\ref{CFmodel}).

In choosing the suitable values of high frequency components $K$, there is not much room for us: 
on one hand $K=2$ is the lowest value meaningful in our analysis due to the same considerations as pointed out in case of the local fits.
On the other hand tests with $K=3$ showed only a small positive effect in the achieved $R^2$ value while significantly increasing the freedom of the
model (distortions) in the region of gaps.
Based on these arguments we decided to fix $K=2$. Total number of parameters in our model is therefore 305, 
which means approximately 10 data points per parameter.

After the optimal parameter values for the model have been fixed the goodness of fit can be further increased by
removing the $3\sigma$ outliers to the initial fit from the data and then refitting again. 
The outliers can be either observationally unreliable points or possible
flares such as that one around April 2000 or HJD 2451650. In our case total of 22 outliers
were detected. 
Removal of these leads to approximately 3\% increase in $R^2$ for global fit.
Therefore in all subsequent CF analysis (for global as well as for local) 
we have used the data set with outliers eliminated.

\subsection{Visualization}\label{CFV}
For visualizing the CF model we use the same technique as introduced
in \citet[p. 4, sect 2.6]{pelt2011}. Firstly, we divide the whole time
span into number of bins with the length of the chosen carrier
period. Secondly, for each bin we normalize the signal amplitude into range
$[-1, 1]$ and then plot it with the corresponding time moment of the
bin and the phase relative to the carrier period. This normalization step
is useful for making the phase behavior of the signal comparable
over the whole time span. Without normalization the features during
high amplitudes will dominate the picture. Here we use both approaches
for the purpose of obtaining more information about the processes
governing the star. At the bottom of the plot we include a so-called
`bar-code' to give information of the density of the data points around the
given time moments. Black indicates densely spaced data, while yellow
indicates sparsely spaced or no data at all. Some previous examples 
using the given technique can be found in
\citep{hackman2013,lindborg2013}.

\subsection{Minima detection, error and significance estimation}\label{MD}
Besides using CF method for visualizing the results, we determine
primary and secondary minima from the model of global fit and compare them to
results obtained from DI analysis and the earlier analysis of a shorter segment of the same data used here with a different method (CPS; \citep{lehtinen2012}.
Error estimates for the minima are
calculated by generating 1000 bootstrap samples from the original data (by reshuffling the residuals of the data points to the initial CF model allowing recurrences), 
repeating the CF analysis for each new data set, and finally, obtaining
the distributions for the minima. We mark a minimum as being
reliable if and only if the following two conditions are satisfied for 
distributions both in time and magnitude: the Kolmogorov-Smirnov test with preassigned significance level 0.01 against a normal distribution 
must pass and the bias of the mean of the distribution from the original estimate 
should be less than the standard deviation of the distribution. 
The error and significance estimates for each minimum are given in the electronic material containing the full global fit.

\section{Methods and results for searching the optimal carrier frequency}\label{carrier}

The selection of the optimal carrier frequency to analyze LQ Hya light curves is far from a trivial task.
There is a wide range of different periods obtained from different data sets and by using different methods.
This is why we need a thorough analysis of the periods to proceed with the CF method.

Taking off from the solar analogy, one might hypothesize that there are at                                                                                                                                 
least four types of periodicities that one needs to deal with:
\begin{itemize}
\item[1.] Behind all the observable activity is the stellar surface rotation                                                                                                                               
and its non-uniformities. In the solar case, the sunspots rather closely                                                                                                                                   
follow the motion of the solar surface seen from Dopplerograms; the only                                                                                                                                   
exception are the longitudinal activity nests that have been reported to show                                                                                                                              
motions that differ from the general pattern. In the absence of asteroseismic                                                                                                                              
data, it is impossible to distinguish between the motion of the plasma and                                                                                                                                 
the spots themselves, and the determination of the rotation period is out                                                                                                                                  
of the scope of this study.                                                                                                                                                
\item[2.] Through photometry, one can hope
at least to be able to determine the mean rotation period of the spots
on the stellar surface, analogous to the Carrington rotation period of the Sun.
\item[3.] Using the DI and ZDI results of significant spot activity in
the low-latitude regions and solar-like rotation law of LQ Hya as                                                                                                                                          
valid assumptions one may also be able to determine the spot rotation
period near the equatorial region, as this should arguably be the
shortest period of rotational origin seen in the periodograms.
\item[4.] Analogously
to the periods derived for the longitudinal activity nest on the solar surface
for the azimuthal dynamo waves on some more evolved rapid rotators,
photometric studies can be used to determine whether any longitudinal
clustering occurs and does the period of the active longitudes
differ from the mean rotation period of the spot distribution.     
\end{itemize}

\subsection{Least-squares fitting}\label{LSF}

For irregularly spaced data sets the most common procedure of frequency analysis
is a simple least-squares fit of a harmonic waveform into the data with the range 
of some trial periods. The particular computational schemes and descriptive statistics vary 
\citep[see e.g.][]{barning1963,lomb1976,scargle1982}. In our analysis we chose the simplest statistic which measures
the relevance of the harmonic under discussion, namely its amplitude. 
\begin{figure}
\begin{center}
\includegraphics[width=\columnwidth]{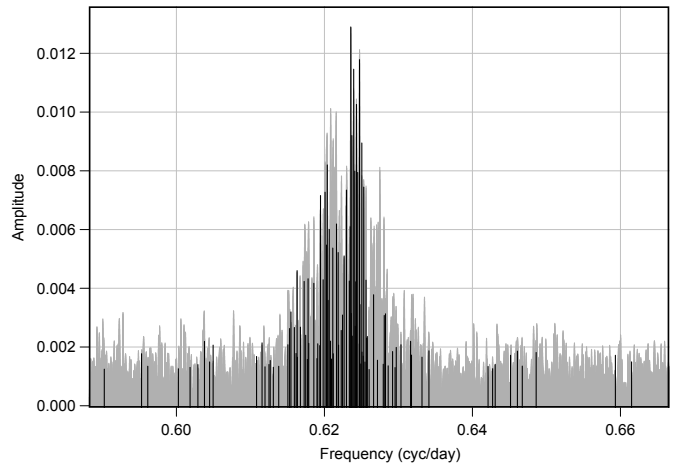}
\caption{Amplitude spectrum for the important period range 1.5-1.7 days (gray) and first 100 frequencies obtained by sequential prewhitening procedure (black).}\label{spectrum}
\end{center}
\end{figure}
\begin{figure}
\begin{center}
\includegraphics[width=\columnwidth]{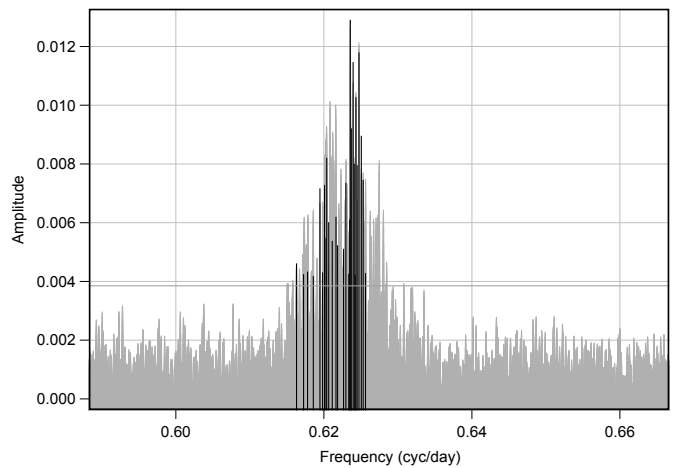}
\caption{The set of 28 strongest amplitudes (black) and original amplitude spectrum (gray).
The horizontal line marks the cut-off level obtained from 100 randomized samples.
}\label{cleaned}
\end{center}
\end{figure}

Before computing the amplitude spectrum, we removed the seasonal means (de-trending) of the full LQ Hya V-band photometry data.
The complexity of the spectrum, depicted in Fig.~\ref{spectrum} with gray color, is obvious and it is very hard to single out any prominent peak from the forest of peaks in it.
The analysis can and must be improved by removing the spurious periods
rising from the gapped nature of the data. For this, we used the so
called pre-whitening method \citep[for a recent application of the
method, see e.g.][]{reinhold2013}. We iteratively removed most significant 
harmonics with estimated amplitudes,
and proceeded in the next step with the least squares fit residuals.
In this way we computed a set of 100 strongest amplitudes which are depicted in Fig.~\ref{spectrum} with black color.
These are not, with high probability, aliases due to the most prominent yearly gap structure with frequency offsets of $\Delta
\nu \approx 1 / 365$.
Even after this cleaning procedure, we still have a very complex set of different frequencies.

A large part of the estimated frequencies with corresponding amplitudes
are well below the noise level and we can disregard them.
To estimate a suitable cut off level, below which the peaks are considered
insignificant, we proceeded in the following way.
We built artificial data sets from the original by reshuffling 
them in time. For every new data set we found its strongest peak. The amplitude cut off level was then chosen as a value of maximum amplitude
in 100 different such random runs. This criterion is rather conservative and we can be quite confident that the 28 periods,
shown with black lines in Fig.~\ref{cleaned},
whose amplitudes occurred to be higher than the cut off level (0.00385; indicated with a horizontal line in Fig.~\ref{cleaned}) 
are not the results of random fluctuations.
As seen from the plot, the set of the selected peaks is now much more localized. Among the peaks are practically all the periods which have been proposed
so far by different authors for different data sets, see Table~\ref{periods}. 
\begin{table}
\begin{tabular}{lcc} \hline
Period & Previous estimates & Source\\
1.601279 & 1.601136 & \citep{jetsu1993} \\
         & 1.601052 & \citep{berdyugina2002} \\
1.600662 & 1.600881 & \citep{strassmeier1997} \\
         & 1.600834 & \citep{berdyugina2002} \\
         & 1.60066  & \citep{kovari2004} \\
1.603893 & 1.6042 & \citep{messina2003} \\
         & 1.60369 & \citep{lehtinen2012} \\ \hline
\end{tabular}
\caption{Periods with strongest peaks in the spectrum compared to previous estimates}\label{periods}
\end{table}

The selected set of periods lay in the interval $1.598406 \dots 1.622572$ days (or in frequency terms $0.6163053\dots 0.6256234$ cycles per day). 
The center of the full range interval is at $\nu_0^{\rm LS} = 0.6209644 \ (P_0^{\rm LS} = 1{\fd}6103984)$ and this is the first logical candidate for the carrier frequency.
The logic behind this choice would be obvious - the full frequency range will be covered on equal grounds. 

Several authors have already reported on the period variability of LQ Hya.
By using local fits, e.g. \citet{messina2003} give period in the range
$1{\fd}5938-1{\fd}6154$ and \citet{you2007} report $1{\fd}60094-1{\fd}60918$
range 
from their analysis.
The obtained range of periods is also in good agreement with a set of
rotation periods obtained from a simple spot modeling procedure by
\citet{alekseev2005}. 
The time span for our main homogeneous data set (D2) is around $9642$
days. Correspondingly the range of rotation counts for the significant
periods is $5942\dots 6032$ with a difference of $90$ rotations. For the
carrier frequency around the center of the estimated range this allows
up to 45 full phase cycle runs in both directions. 
In Sect.~\ref{CFSFP}, however, we concluded that the optimal harmonic count for low
frequency modulation curve is around $30$ cycles. 
From this follows
that in principle the very sharp (momentous) frequency jumps from one
side of the range to the other can become smoothed
out to some extent.

On the other hand, the second longest seasonal gaps in the data is around
$192$ days and corresponding cycle counts are $199\dots 120$
Obviously, for the carrier periods in the middle of the
full range the phase migration during the gaps is certainly less than
a full turn, cf. Eq.~(\ref{cyclecount}).
For the longest
gap in the data (around 300 days), however, we can theoretically have a cycle
count error and consequently the approximated solution in this region
can not be regarded reliable.

Finally, one interesting observation that can be seen in
Fig.~\ref{cleaned} is that the distribution of the significant frequencies
is bimodal, i.e. there are two bunches of them. Interestingly enough, as
seen from Table~\ref{periods} typical solutions are
concentrated in the rightmost bunch of periods,
the cut-off being very sharp on the higher frequency (shorter period) side. 

The bimodal structure of the period distribution leads us to another
trial hypothesis -- that we have here a case when a certain frequency in
between the two bunches is more or less periodically modulated (with
period around 2000-3000 days). This hypothesis was already set up in
an earlier paper \citep[see][]{berdyugina2002}. To check it once
again we carried out the corresponding analysis for our significantly longer data set D2.

\subsection{Carrier from a multiperiodic model}\label{TwoModulators}
    
The simplest conceivable model for the slowly modulated signal is a time series which
depends on the carrier period $P_{0}^{\rm MP}=1/\nu_{0}^{\rm MP}$ and modulating period $P_{\rm mod}=1/\nu_{\rm mod}$ in a coupled way. That means that
all positive combination frequencies
\begin{equation}
\nu_{i,j}=i\times\nu_{0}^{\rm MP}+j\times\nu_{\rm mod},i,j=-N,\dots,N,
\end{equation}
can take part in waveforming. In the simplest case of $N=2$ there will essentially be 25 different trigonometric
terms (constant included) which must be fitted into the observed data using the least squares method \citep[see][]{berdyugina2002}.
The periods producing the best fit will then used to build final model.

\begin{figure}
\begin{center}
\includegraphics[width=\columnwidth]{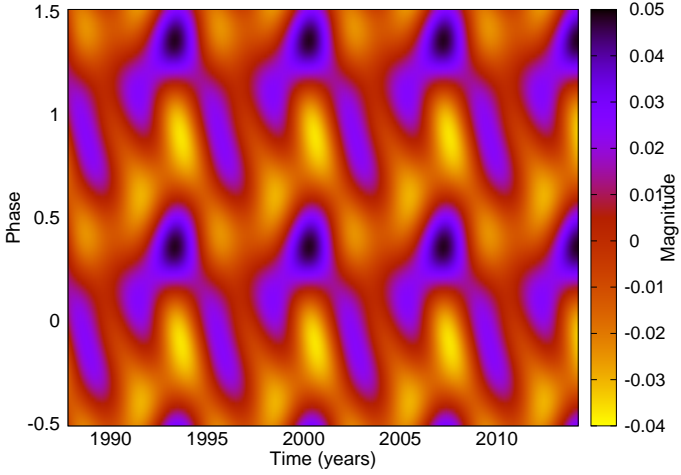}
\caption{Folded and stacked light curve model with periods $P_{0}^{\rm MP} = 1{\fd}602680$ and $P_{m} = 2534{\fd}6$. This regular structure helps to
describe only $R^2 = 23.542\%$ of overall variability.}\label{regularfit}
\end{center}
\end{figure}

We performed this kind of analysis for the data set D2. The resulting pair of periods
occurred to be $P_{0}^{\rm MP} = 1{\fd}602680\pm 0{\fd}0000027$ and $P_{\rm mod} = 2534{\fd}6 \pm 6{\fd}6$ (roughly 6.94 years).
The resulting fit can be visualized by the same devices as the carrier fit. We divide the solution into period length fragments
and stack them properly. In Fig.~\ref{regularfit} our solution is depicted. The $R^2$ value for the least squares fit for
these two periods is quite low $(23.6\%)$. However, it is remarkable that the modulation period is very similar to
values often quoted as the long cycle length of LQ Hya \citep[e.g.][]{jetsu1993,olah2000,berdyugina2002,alekseev2005}.

\begin{figure}
\begin{center}
\includegraphics[width=\columnwidth]{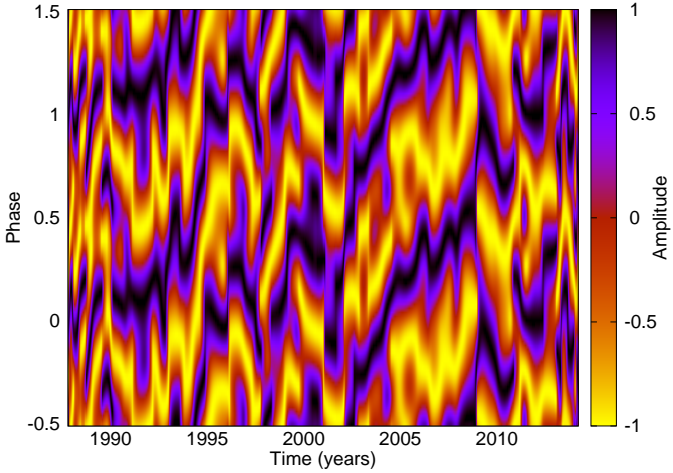}
\caption{Carrier fit obtained with the carrier $P_{0}^{\rm MP} = 1{\fd}60268$ obtained from the two-periodic model of Sect.~\ref{TwoModulators}}\label{carrierfit}.
\end{center}
\end{figure}

If we now base our consideration on the idea that the possible doubly periodic component is a relevant aspect of the overall variability 
then we can use the
computed carrier value as a model based carrier $P_{0}^{\rm MP}$ for the CF procedure. The result of this analysis is shown in Fig.~\ref{carrierfit}.
Unfortunately, the regular structure of the simple model is largely lost and a number of other variability elements dominate the picture.


\subsection{Carrier from the phase dispersion analysis}\label{D2}

From Fig.~\ref{cleaned} we can well see that 
the period computed as a mean from the formal range of significant periods is not very representative due to the bimodal nature of the distribution.
This is also true if the peak with maximum amplitude was selected as a carrier.
Slightly better method would be the one used in \citet{lehtinen2012} where the best period was selected using phase distributions
of the light curve extrema. However in this case the interplay between different frequencies can shift minima or maxima and
obscure the general picture. 

\begin{figure}
\begin{center}
\includegraphics[width=\columnwidth]{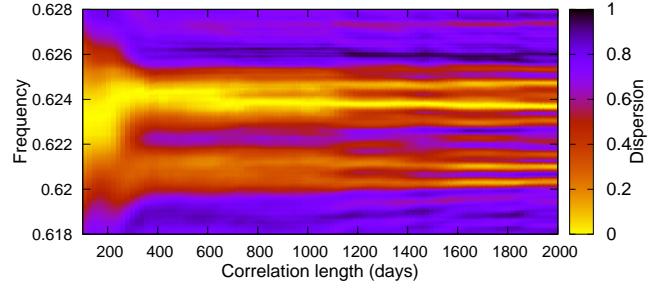}
\caption{Phase dispersion statistic $D^2(P)$ for a range of correlation lengths. 
The dispersion spectrum becomes strongly asymmetric around 230 days and then splits into set of peaks.}\label{phasedisp}
\end{center}
\end{figure}

To our understanding, the best method for computing the mean period of spots comes from phase dispersion analysis \citep{pelt1983,lindborg2013}. It is based on the following simple statistic: 
\begin{equation}
D^2(P) = \frac{1}{2\sigma^2}\frac{{\sum\limits_{i = 1}^{N - 1} {\sum\limits_{j = i + 1}^N {g(t_i } } ,t_j ,P,\Delta t)[f(t_i ) - f(t_j )]^2 }}{{\sum\limits_{i = 1}^{N - 1} {\sum\limits_{j = i + 1}^N {g(t_i } } ,t_j ,P,\Delta t)}},
\end{equation}
where $f(t_i),i = 1,\dots,N$ is the input time series, $\sigma^2$ is its variance, $g(t_i,t_j,P,\Delta t)$ is the selection function  which is significantly greater than zero only when
\begin{eqnarray}
t_j  - t_i  &\approx& kP,k =  \pm 1, \pm 2, \ldots {\rm \ \ \ and}\\
\left| {t_j  - t_i } \right| &\le& \Delta t.
\end{eqnarray}  
In the latter condition $\Delta t$ is so called {\it correlation length}. For the particular case when $\Delta t$ is longer than the full data span
the $D^2(P)$ statistic is essentially a slight reformulation of the well known Stellingwerf statistic \citep{stellingwerf}. 
As the correlation length is made shorter, we match nearby cycles in a progressively narrower region, and consequently estimate a certain mean period, 
which needs not to be coherent for the full time span. This is well illustrated in Fig.~\ref{phasedisp}, where we show the $D^2(P)$ statistic for the range
of trial frequencies as function of the correlation length, $\Delta t$, using color contours.
We see that for a small enough correlation length the $D^2(P)$ statistic yields a
rather symmetric single minimum. Above that limit the frequency spectrum starts to
distort and eventually splits into separate branches. 
Finally at large correlation lengths we obtain a forest of minima similar to the results presented in Sect.~\ref{LSF}.
We interpret this behavior in the following way: for short coherence
times, the periodogram is dominated by the mean pattern of spot
motions, while at longer coherence times the signatures of more
persistent spot structures, the rotation of which differs from the
mean spot flow, take over and give the strongest signal.

\begin{figure}
\begin{center}
\includegraphics[width=0.4\textwidth]{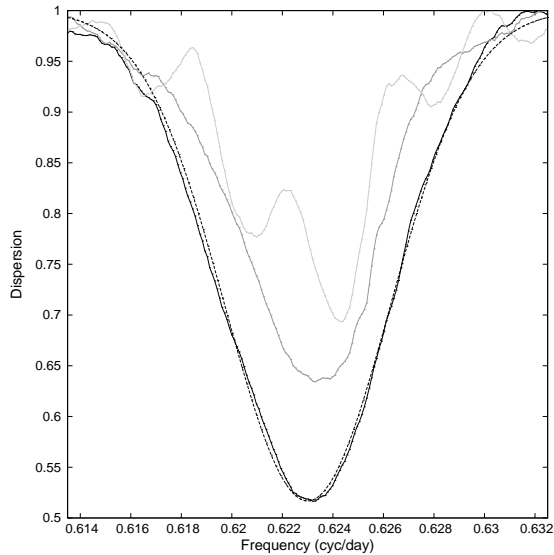}
\caption{Curves of the phase dispersion statistic $D^2(P)$ for different correlation lengths in days: 100 (black), 200 (gray) and 300 (light gray). 
Dashed curve corresponds to the best-fitting Gaussian to the curve with correlation length 100 days.}\label{phasedisp2}
\end{center}
\end{figure}

Our aim is to
determine the limiting correlation length at which a single minimum is still obtained and use the corresponding period of the phase
dispersion minimum as a plausible carrier period. 
Here we must point out two
problems: firstly the minimum of $D^2(P)$ statistic is usually very wide, and
secondly the shape of the peak somewhat deviates from
symmetric form (see for details Fig.~\ref{phasedisp2}). 
This prevents us from determining a sufficiently accurate value of the minimum directly from the curve of the statistic.
Instead of that we fit Gaussian profile to the curve of $D^2(P)$ statistic. Our task is to estimate the free parameters
of this gaussian curve (mean $\mu$ and variance $\sigma^2$) for which the distance to the curve of the $D^2(P)$ statistic is minimal. 
The value of the mean obtained this way represents the optimal carrier frequency and the variance represents the scatter of the periods around it.
From our analysis it turns out that with the correlation length of 100 days the curve of the $D^2(P)$ statistic is singular and still symmetric enough,
due to which we choose this correlation length as the limiting one, and determine the mean period from this curve. The best-fitting Gaussian has the mean $\mu=0.62300$ and 
standard deviation $\sigma=0.00325$ (both in cycles per day).
In the time domain the corresponding values are $P_0^{\rm D2}=1{\fd}60514$ and $\Delta{P_0^{\rm D2}} \approx 0{\fd}0084$. 
For comparison, quite a similar value,$P_{\rm w}=1{\fd}6043$, was obtained by \citet{lehtinen2012} 
by calculating the weighted mean of the periods determined for independent subsets of the whole data.

In the current study we estimated the significance of the 
mean cycle length $P_0^{\rm D2}$ by testing the null hypothesis that the peak of the minimum is drawn
from the distribution of random fluctuations. For that purpose we generated 1000 samples from the 
original data set via reshuffling the measured magnitudes and then calculated the value of the $D^2(P)$ 
statistic for each new data set using correlation length of 100 days. In such a way we 
obtained a distribution for the minimum of the $D^2(P)$ statistic caused by random fluctuations. 
The results showed that in our case the null hypothesis can be rejected with preassigned 
significance level of 0.05 as the minimum of the $D^2(P)$ for the original data remains well below the 
995th value from the sample distribution which was located around 0.97.

We believe that the carrier value obtained by using the $D^2(P)$ statistic, which is computed for short correlation lengths,
is well grounded. The other values (the strongest peak in the spectrum, central value of the range, mean value of the set of periods, periods based on
extrema) tend to bring in an amount of contingency,
as using any of them would in practice mean choosing a certain locally active period for the CF analysis of a global nature.
  
\section{CF results}\label{res}
In the following we will use the obtained mean cycle length period of the spots $P_0^{\rm D2}$
as the carrier period, and compute a global as well as local fits based on it.

\subsection{Global fit}\label{resGRref}
\begin{figure*}
\begin{center}
\begin{tabular}{c}
\includegraphics[width=0.48\textwidth]{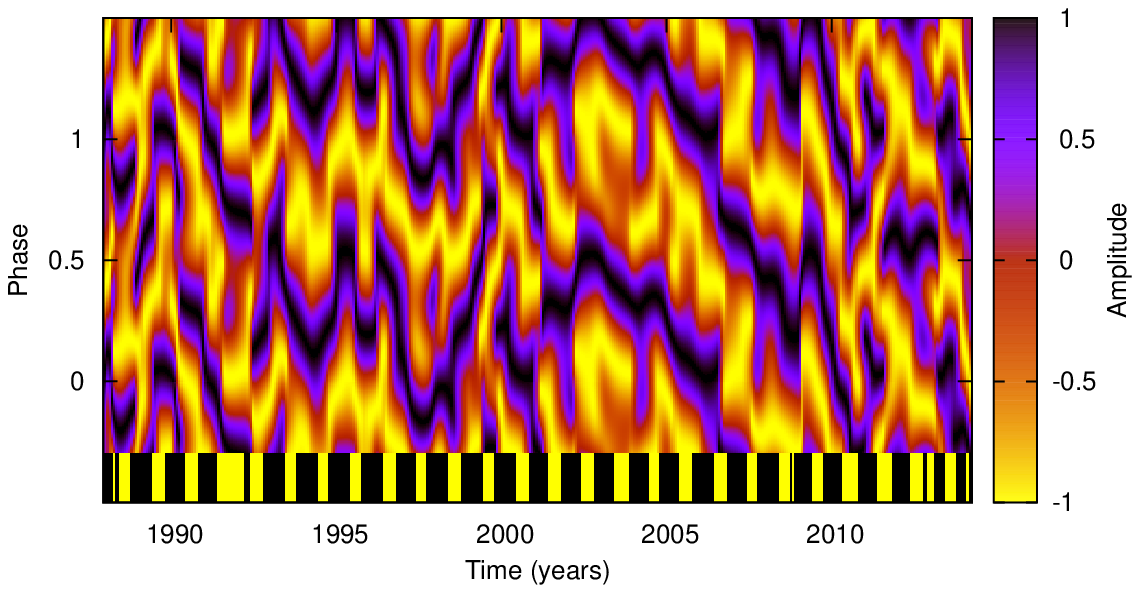}
\includegraphics[width=0.48\textwidth]{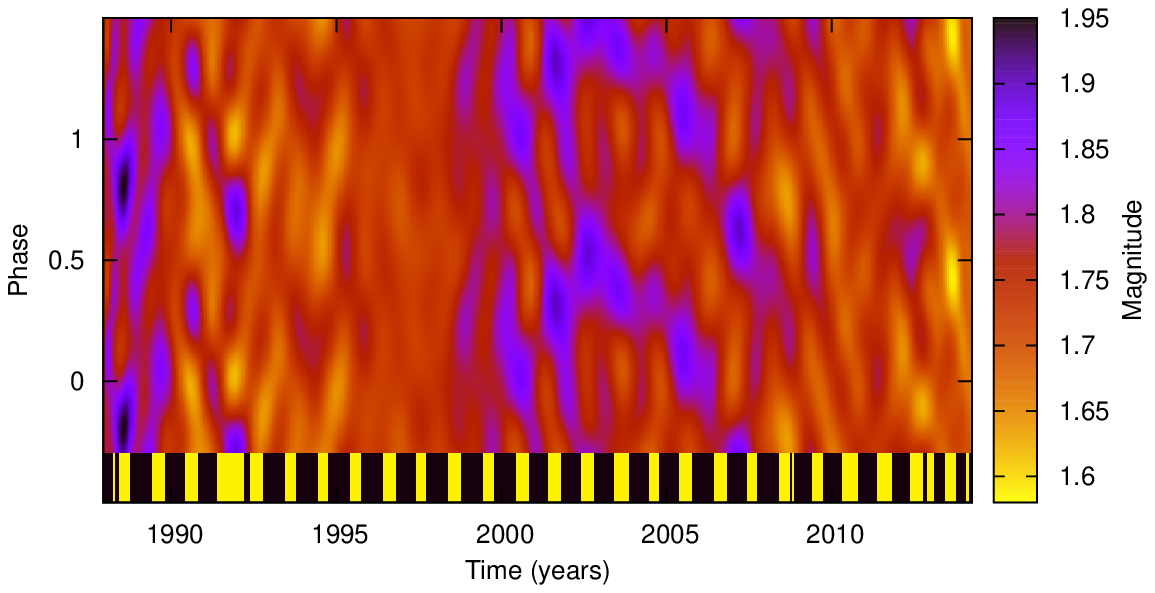}
\end{tabular}
\caption{Phase diagram for global fit with carrier period $P_0^{\rm D2}=$1{\fd}60514, K=2 and L=30. 
On left with normalized amplitudes, on right with actual magnitudes}\label{refinedGF}
\end{center}
\end{figure*}

We perform a global CF analysis with the selected parameters$P_0^{\rm D2}$, $K=2$, $L=30$, and $C=1.2$, which
gives a model with $R^2=91.7 \%$.
Consequently, the carrier fit model describes a rather large part of the overall input data variability.
The resulting phase diagram is displayed in Fig.~\ref{refinedGF}, wherefrom it is evident that the
phase behavior is characterized by up- and downward trends, while epochs during which the minima would occur
at constant phases are almost totally absent. 
On one hand the slopes of the trends correspond to the different periods, on the other hand the lengths and the locations of
the trends give hints of the "duty times" of these cycles (in other words when and for how long these periods dominate).
The most noticeable features are two downward trends during the years 2005 -- 2008 and 1991 -- 1993. These correspond to the periods 
1${\fd}$6037 and 1${\fd}$6026 days accordingly, which belong to the strongest peaks found previously by frequency analysis. 
During other epochs the phase behavior is changing more rapidly in time while the abrupt phase shifts 
seem to appear over the whole time span, even during the strongest trends.

\subsection{Local fits}\label{resLF}
In the current study local analysis of the data is carried out mainly for the purpose of visualizing the phase behavior of
the signal in more detail than it is possible with the global analysis. 
Due to insufficient number of modulators in the global fit, the whole spectrum of the data is not covered. This leads to the smoothing of the real signal on small time scales. To get better results we perform a local CF analysis similarly as it was done in 
\citet{hackman2013} and \citet{lindborg2013}.
Segments contain relatively low number of data points and in some cases considerable gaps are present, thus
the maximum allowed number of modulators is significantly smaller than
that one for the global fit. Therefore one could assume that no higher precision in the results is achievable.
However, the reality is exactly the opposite, we obtain better coverage of the spectrum due to 2 reasons:
firstly the spectrum of each segment separately is narrower than that one of the whole data set and 
secondly the frequencies of the modulators in the model are much higher than the ones used in case of the global fit.

\begin{figure*}
\begin{center}
\includegraphics[width=\textwidth]{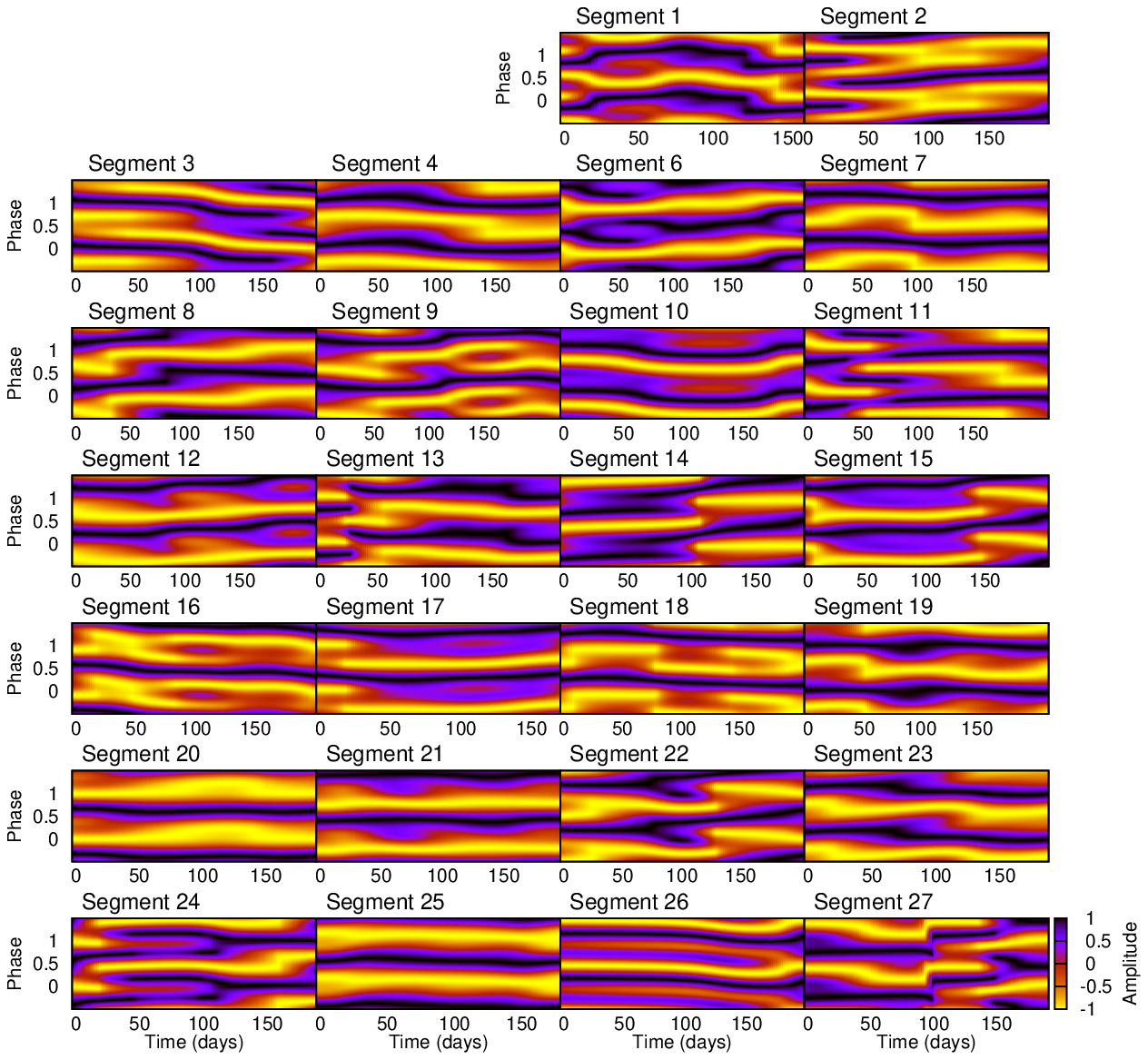}
\caption{Phase diagrams of local fits with refined carrier period $P_0^{\rm D2}=$1{\fd}60514}\label{localfits}
\end{center}
\end{figure*}

Splitting the data into segments suitable for the local analysis is done using the following rules: we start with the earliest data point and calculate the 
difference in time between pairs of data points next to each other $d_i=t_i-t_{i-1}$. If $d_i<100d$ the data point at $t_i$ is 
added to the segment, otherwise a new segment is started and the process is repeated. Using the above algorithm we obtained 27
segments, the details of which are given in Table~\ref{segments}. CF-s for each segment were applied with the period of the mean cycle length 
$P_0^{\rm D2}=1{\fd}60514$.
The number of harmonics $K$ and modulators $L$ optimal for each segment were determined by finding the minimum for the BIC. 
For all segments $K=2$ turned out to be suitable except for the segment 5, which had only 27 points so that realistic modeling was impossible. Due to the cycle count problem $L$ was further decreased for the segments 1 and 26. 
Exact number of parameters $L$ used in the model and the $R^2$ value achieved for each segment are summarized in Table~\ref{segments}. 
We can see that for most of the segments the goodness of fit is even higher than 90\%, but for segments 3, 10, 11 ands 26 it is quite low. In case
of the segment 26 this can be explained by a big gap in the data and the low number of points. The other above mentioned three segments are, 
however, quite densely populated. 
This might be an indication that the signal is more complex in these segments than what is possible to model with given number of data points.
The resulting phase diagrams for all 27 segments are shown in Fig.~\ref{localfits}, where the primary minima can be found by following 
black or dark blue features, while secondary minima appear either as red features between yellow features, or violet features between red features. 

In Table~\ref{segments} occurrences of flip-flops are marked with the '+' symbol, totaling to 4 events in segments 2, 11, 13 and 24. Other type of
disrupted phase behavior events are marked with '?', constituting the segments 1, 6, 8, 9, 12, 15, 22, 24 and 27. These are either single phase jumps of primary minima or swaps between primary and secondary minima 
less than 0.5 in phase. Clear upward trends can be seen in segments 6, 12 14, 15, 22 and a relatively gentle downward trend in segment 16. These are marked with the 
symbol '/' in the same table. 

A similar analysis was carried out by \citet{lehtinen2012} using the CPS method for the same data set, except for the last three seasons.
The results of this study are in agreement with ours -- most of the interesting features can be seen on the phase plots from both studies. 
Some differences occur for the segments 3 and 6 
where some of the minima detected by CF are absent in the case of CPS. For segment 19 there is no secondary 
minima from CPS, but it can be seen during the first 20 days in case of CF. No comparison between the results is available for segments 4 and 5 due to 
low number of data points.

\begin{figure}
\begin{center}
\includegraphics[width=\columnwidth]{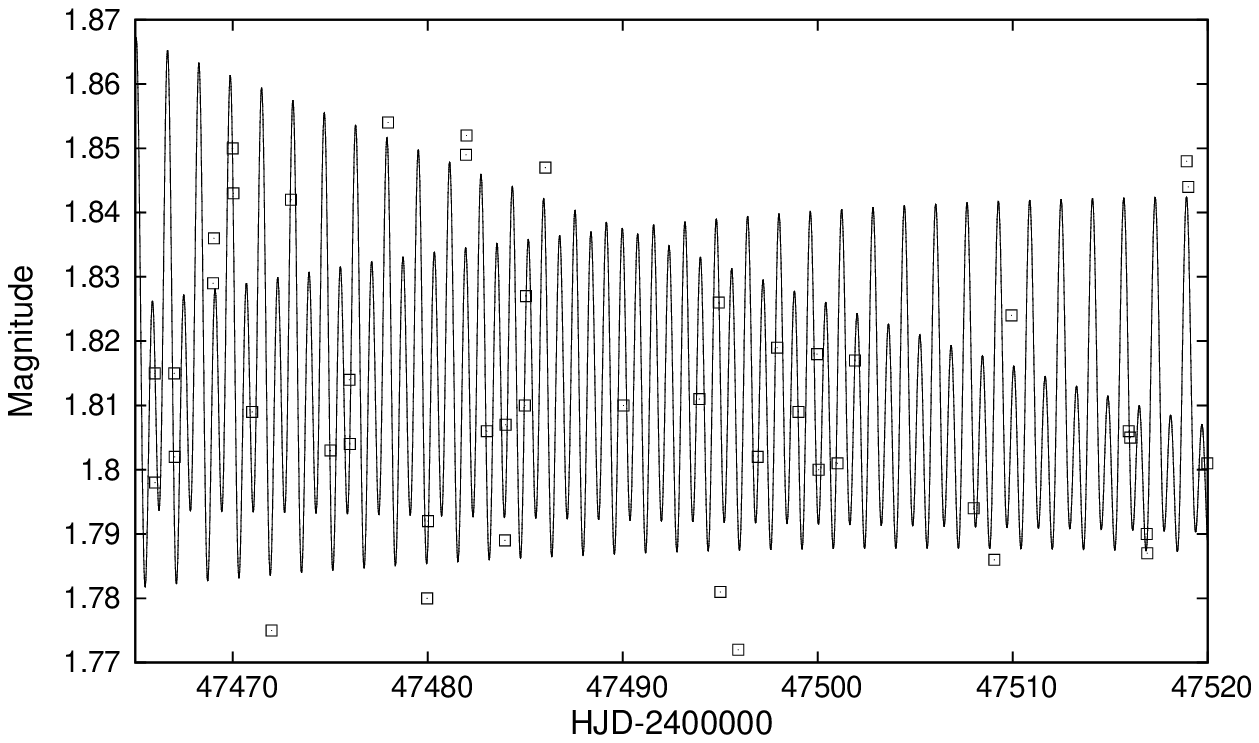}
\caption{Zoom-in to the global CF near the flip-flop event detected in the segment 2. The data points are drawn as small rectangles.}\label{flipflop}
\end{center}
\end{figure}

We have calculated epochs of possible flip-flop events also from the global fit using the definition from
\citet{hackman2013}:
\begin{itemize}
\item the region of main activity shifts about 180 degrees from
the old active longitude and then stays on the new active
longitude or
\item the primary and secondary minima are first separated by about 180 degrees, after which the secondary
minimum evolves into a long-lived primary minimum, and
vice versa.
\end{itemize}
Two additional restrictions were added to the above scheme: firstly, we counted only those 
events for which the phase shift lies between 0.45 and 0.5; secondly, the primary and secondary minima at the moment of flip-flop must be reliable
according to the error estimates from bootstrap runs.
The results show that four of the total six flip-flop events detected from global CF reside within the data, while two occur in gaps. Moreover, three of these
are located in the segments 2, 13 and 24 for which we have detected flip-flops also from local fits.
One of the flip-flops, namely within segment 11, was detectable only from the local fit.
The epochs of all 7 detected events are depicted in Fig.~\ref{minima} with thick green vertical lines. Example of the global CF model around the
flip-flop event in segment 2 can be seen in Fig.~\ref{flipflop}. Following the light curve from left to right we notice that the magnitude of the primary minima decreases while that one of the secondary minima increases. After around HJD 2447490 both minima "swap" their magnitudes, the change in phase corresponding to 0.5.

In \citet{berdyugina2002} a 5.2 year flip-flop cycle was reported. In the light of our current study this periodicity cannot be confirmed:
on one hand this is due to the small number of events detected and on the other hand
some of the detected events are separated only by 2 or 3 years.

\begin{table*}
\begin{center}
\begin{tabularx}{425pt}{lccccccc} \hline
Segment & $HJD-2400000$ (Date) & $\Delta{T_{seg}}$ & $\Delta{T_{gap}}$ & $N$ & $L$ & $R^2$ & Events \\ \hline
1 & 47141 (1987-12-11) - 47304 (1988-05-22) & 164 & 47 & 42 & 2 & 96\% & ? \\ 
2 & 47460 (1988-10-25) - 47660 (1989-05-13) & 201 & 8 & 215 & 1 & 80\% & + \\ 
3 & 47832 (1989-11-01) - 48027 (1990-05-15) & 196 & 13 & 166 & 1 & 65\% & - \\ 
4 & 48189 (1990-10-24) - 48394 (1991-05-17) & 206 & 39 & 46 & 1 & 93\% & - \\ 
5 & 48696 (1992-03-14) - 48759 (1992-05-16) & 64 & 14 & 27 & NA & NA & NA \\ 
6 & 48911 (1992-10-15) - 49132 (1993-05-24) & 222 & 19 & 88 & 2 & 89\% & ?, / \\ 
7 & 49277 (1993-10-16) - 49499 (1994-05-26) & 223 & 8 & 137 & 2 & 96\% & - \\ 
8 & 49645 (1994-10-19) - 49866 (1995-05-28) & 222 & 14 & 129 & 2 & 96\% & ? \\ 
9 & 50006 (1995-10-15) - 50226 (1996-05-22) & 221 & 11 & 152 & 2 & 84\% & ? \\ 
10 & 50391 (1996-11-03) - 50595 (1997-05-26) & 205 & 12 & 148 & 1 & 65\% & - \\ 
11 & 50736 (1997-10-14) - 50955 (1998-05-21) & 220 & 13 & 134 & 1 & 59\% & + \\ 
12 & 51103 (1998-10-16) - 51325 (1999-05-26) & 223 & 8 & 188 & 2 & 85\% & ?, / \\ 
13 & 51474 (1999-10-22) - 51687 (2000-05-22) & 214 & 10 & 126 & 3 & 95\% & + \\ 
14 & 51861 (2000-11-12) - 52052 (2001-05-22) & 192 & 11 & 73 & 1 & 92\% & ?, / \\ 
15 & 52214 (2001-10-31) - 52421 (2002-05-26) & 208 & 19 & 81 & 2 & 93\% & ?, / \\ 
16 & 52582 (2002-11-03) - 52785 (2003-05-25) & 204 & 12 & 100 & 3 & 97\% & -/ \\ 
17 & 52977 (2003-12-03) - 53149 (2004-05-23) & 173 & 9 & 85 & 2 & 94\% & - \\ 
18 & 53299 (2004-10-20) - 53506 (2005-05-15) & 208 & 21 & 92 & 2 & 94\% & - \\ 
19 & 53660 (2005-10-16) - 53876 (2006-05-20) & 217 & 10 & 107 & 2 & 97\% & - \\ 
20 & 54044 (2006-11-04) - 54238 (2007-05-17) & 195 & 8 & 101 & 2 & 99\% & - \\ 
21 & 54400 (2007-10-26) - 54599 (2008-05-12) & 200 & 13 & 124 & 3 & 98\% & - \\ 
22 & 54761 (2008-10-21) - 54966 (2009-05-14) & 206 & 25 & 70 & 2 & 96\% & ?, / \\ 
23 & 55121 (2009-10-16) - 55312 (2010-04-25) & 192 & 15 & 97 & 2 & 95\% & - \\ 
24 & 55499 (2010-10-29) - 55690 (2011-05-08) & 191 & 6 & 127 & 2 & 90\% & + \\ 
25 & 55875 (2011-11-09) - 56054 (2012-05-06) & 179 & 12 & 116 & 3 & 99\% & - \\ 
26 & 56226 (2012-10-25) - 56426 (2013-05-13) & 200 & 95 & 44 & 1 & 77\% & ? \\ 
27 & 56589 (2013-10-23) - 56783 (2014-05-05) & 194 & 60 & 70 & 2 & 94\% & ? \\ \hline
\end{tabularx}
\end{center}
\caption{Summary of the local CF analysis results. Columns from left to right: 
segment number, start and end time epochs for the segment in HJD - 2400000 and corresponding dates, 
length of the segment in days $\Delta{T_{seg}}$, length of the longest gap $\Delta{T_{gap}}$ in days, number of observations $N$, 
number of modulators $L$, goodness of fit $R^2$ and the type of the event that could be detected in 
the segment (if any). NA stands for the segment having not enough data points for meaningful CF analysis, '-' for smooth phase behavior, 
'+' for disrupted phase behavior of flip-flop type, '?' for disrupted phase behavior that can not be associated with a flip-flop type event, '/' phase drifts.}
\label{segments}
\end{table*}

\subsection{Comparison with Doppler Imaging}\label{resDI}

\begin{figure*}
\begin{center}
\includegraphics[width=\textwidth]{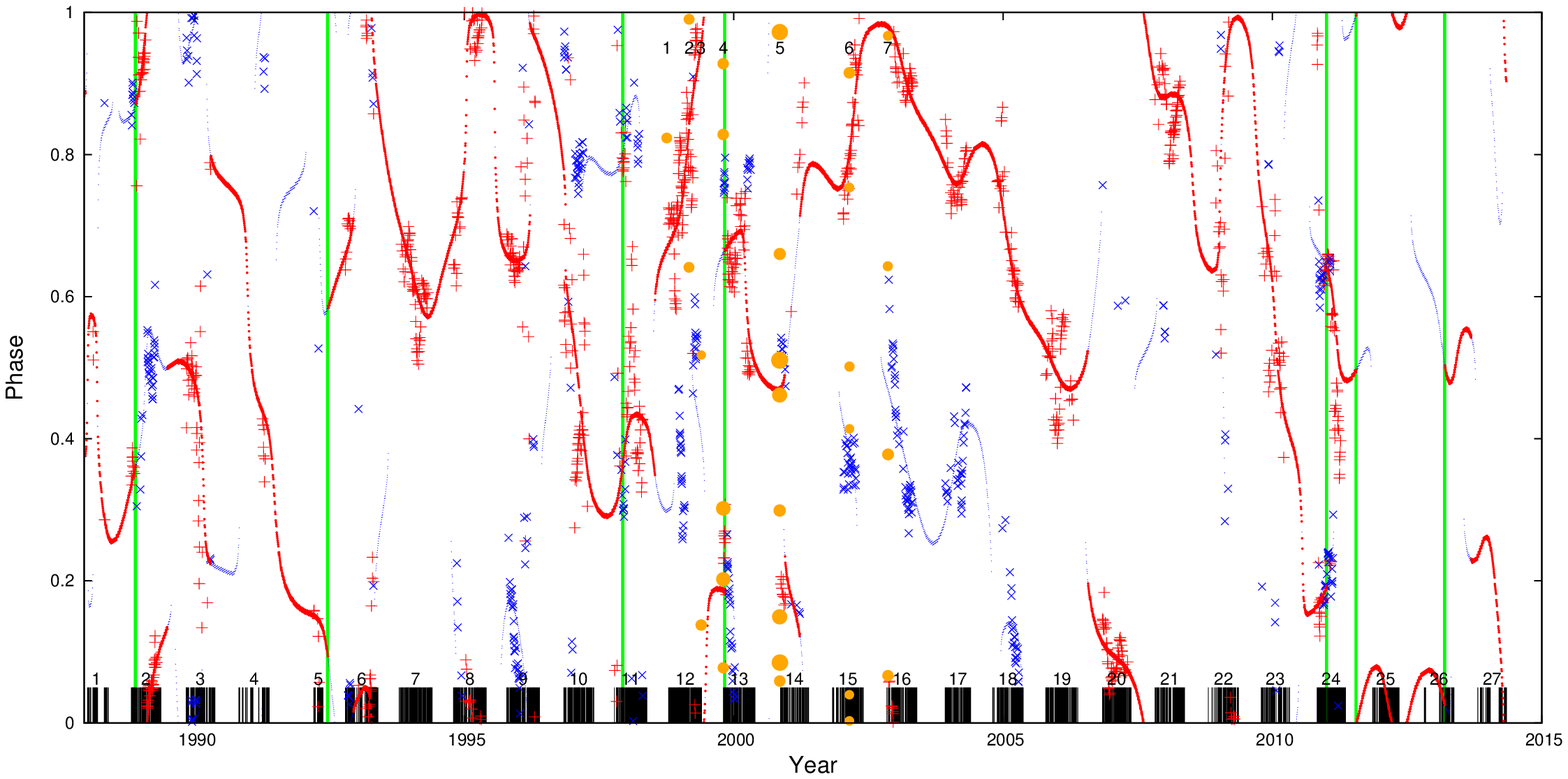}
\caption{Phases of the minima for period $P_0^{\rm D2}$=1{\fd}60514. Red dots: primary minima from CF, red pluses: primary minima from CPS, 
blue points: secondary minima from CF, blue crosses: secondary minima from CPS, orange circles: minima from DI, bold green vertical lines 
denote the epochs of possible flip-flop events. The black bars with the numbers correspond to the photometric observing seasons, 
the numbers on top of the figure mark the observing seasons of DI.}\label{minima}
\end{center}
\end{figure*}

Using our CF model for a global fit with the period of mean cycle length of 1{\fd}60514 we determined the epochs of primary and secondary
minima. In Fig. \ref{minima} these epochs are plotted against the phase
of the same period. Larger red and smaller blue dots represent the locations of primary and secondary 
minima respectively. As noted above,1000 bootstrap samples were calculated for the
global CF to obtain error estimates for the minima. However, for a clearer 
visualization these are omitted from the figure, but are included in the online material, where
the global CF solution is given.

In \citet{cole2014b} the DI technique was applied to spectrometry of 7
observing seasons. The resulting surface temperature maps were used to
determine the epochs of the temperature minima, interpreted as starspots, for every season.
Especially high activity level, i.e. a large amount of cool spots, was observed to occur during October 1999 to
November 2000. In Fig. \ref{minima} we have plotted the retrieved spot epochs with
orange circles, where the size of the circle reflects the temperature of
the given spot (a larger circle corresponds to a lower temperature). On
the same plot we have also included the minima obtained from the CPS
method published in \citet{lehtinen2012}. Red pluses and blue crosses respectively represent the primary and 
secondary minima.

As expected the minima obtained from the global fit serve as the averaged
values for the minima obtained from the CPS. 
Agreement with the results from DI is also quite satisfactory. 
There is a quite good match between DI and other
models for some of the minima found in seasons 3 to 7. 
Even though neither active longitudes nor flip-flop type events were seen in DI, both global and local
CF analysis reveal a possible flip-flop. 
However, it is not reasonable to
search for the full agreement between photometry and DI. For instance the photometry is 
affected by the limb darkening and surface area projection effects of the active regions, 
so that one to one correspondence between the strongest minima in photometry and 
the lowest temperature regions in DI cannot be expected.
We also note that the observing
seasons for spectrometry and photometry mostly do not
overlap. In addition a low S/N ratio as well as less than ideal phase coverage for several 
observing seasons was reported, which increases the uncertainties even further \citep{cole2014b}.

\section{Discussion}\label{disc}
If we make an assumption that the scatter around the mean period is
the result of differential rotation we can use the half-width of the
minimum as an approximation for the standard deviation. Based on this we can
estimate the differential rotation coefficient using the formula by
\citet{jetsu1993}:
\begin{equation}
k = \frac{6 \Delta P_0^{\rm D2}}{P_0^{\rm D2}},
\end{equation} 
where in our case $\Delta P_0^{\rm D2} \approx 0{\fd}0084$ is taken as the $\sigma$ of the 
closest gaussian curve to dispersion statistic with correlation length of
100 days. Substituting the value into the equation leads to $k \approx
0.032$ which is in rough agreement with previous estimates from photometry
\citep[e.g.][]{lehtinen2012,you2007,jetsu1993}, $k$=0.015...0.025, but
significantly larger than the values obtained from DI
\citep[e.g.]{kovari2004,donati2003}, $k$=0.002...0.006.

The Coriolis number is an estimate of the strength of the rotational
influence over turbulent convection, and can be written as
\citep{Saar1999}
\begin{equation}
{\rm Co}=4 \pi \tau_{\rm c}/P_{\rm rot},\label{Co}
\end{equation}
where $P_{\rm rot}$ is the rotation period of the star and $\tau_{\rm
  c}$ is the convective turnover time. \citet{Ossendrijver1997}
presented an extrapolation method to the theoretical calculations of
\citet{kim1996}, to estimate the convective turnover time from the B-V
index. \citet{lehtinen2012} applied this technique and arrived at an
estimate of $\tau_{\rm c} \approx 33.5$\,days for LQ Hya. On the other hand, they
used their CPS method to compute the time scale of change, denoted
with $T_{\rm c}$, for each individual data set investigated, and as an
average of all the analyzed segments, found a value of 50.5\,days.  As
this quantity describes the typical time in which changes in spot
configuration occur, it can be postulated to have some relation to the
convective turnover time.

The Coriolis number of LQ Hya, based on the above stated values of the turnover time, lies within the range
260-400, these numbers being huge in comparison to the Sun with Co of
roughly 6 with the definition Eq.~(\ref{Co}). In the study of
\citet{Saar1999} stars were observed to cluster on certain activity
branches, when their Coriolis number and rotational vs. magnetic cycle
periods, $P_{\rm rot}/P_{\rm cyc}$, were plotted.  LQ Hya was termed
as an anomalous object, falling in the transition region between the
active (A) and superactive (S) branches, based on the values
$\tau_{\rm c} \approx 20.9$\,days by \citet{gunn1998} and $P_{\rm cyc}
\approx 7$ years by \citet{strassmeier1997} used back then. The majority
of magnetic cycle determinations still falling into the range of 6-7
years, together with the usage of the alternative methods to determine
the convective turnover time to compute the Coriolis number, would
place the object to an even more anomalous place in the diagnostic
diagram, clearly further away from
the other stars (although only a few of them identified) in the
transition region.  This makes LQ Hya a very fascinating object to
follow up and study further. These considerations, on the other hand,
might also indicate that the division into active and superactive
branches in the diagnostic diagram is not as meaningful as the
separation into the inactive and active ones (the so-called
Vaughan-Preston gap).

The difference between the retrieved mean rotation period of the star
and the most pronounced coherent phase structure during 2005 and 2008
($P_{\rm coh}=1{\fd}6037$)
is roughly 0.00171 days. During the aforementioned years, the trend in
the phase-time diagram is nearly linear, the spot structure going
faster (nearly linear downward trend) with respect to the mean
rotation of spots. Previously, such behavior has been seen on more
evolved rapid rotators \citep{lindborg2011,hackman2013}, and has been
interpreted as being due to either latitudinal differential rotation or
an azimuthal dynamo wave. The effect is definitely the clearest in the
primary component of the binary system II Peg, where a clear linear
trend is visible for over ten years \citep[see e.g.][]{lindborg2013},
whereas only disrupted up- and downward trends were seen in the single
giant FK Com \citep{hackman2013}. 
The complexity level and the type of phase behavior seen in FK Com are
similar to the ones reported for LQ Hya in the present paper; this
could be an indication of the binarity of II Peg having an influence on
stabilizing the active longitudes in comparison to single stars, LQ
Hya and FK Com representing this class.

In the case of II Peg and FK Com, however, it was difficult to
definitely rule out the differential rotation scenario, as for
instance in the case of II Peg, similar magnitude of drift could have
been caused by an anti-solar differential rotation profile with $k$
comparable to the values deduced from observations.  Also, in the case
of LQ Hya, the situation is somewhat similar: the deduced values of
the differential rotation parameter $k$ range between 0.002 ... 0.032,
the smallest ones being obtained from Doppler imaging, the largest
ones from photometry. 
DI and ZDI studies
\citep[e.g.][]{strassmeier1993,rice1998,donati1999,donati2003,kovari2004,cole2014b}
indicate that the majority of the spot activity of LQ Hya occurs at
two different latitudinal regions, namely at high- and nearly
equatorial regions. It cannot be ruled out that in such a system, 
spots could be drifting from the high-latitude location to the
lower-latitude one, in which case they would gradually attain
faster rotation rate due to the most likely solar-like rotation
law of the object.
The maximal latitude range of the drifting structure versus
the mean spot latitude would be of the order of $\pi/4$, and therefore
the implied $k$ for the spot structures roughly half of the
differential rotation parameter, i.e. $k_{\rm dr}^{\rm exp}$ = 0.001
... 0.016.  The value that can be directly computed from the period
difference of the coherent structure and the mean movement of the
spots reads
\begin{equation}
k_{\rm dr}^{\rm drift} = \frac{P_0^{\rm D2} - P_{\rm coh}}{P_0^{\rm D2}} \approx 1.1 \times 10^{-3},
\end{equation}
which is close to the lower limit of the values derived from Doppler imaging. Therefore, again, there is certainly enough differential rotation
on the object to be the cause of the observed phase-time drift.
One must note that such a drift would not cause strictly linear (but
curved) trends in the phase-time plots \citep[see][for a simulated
example]{pelt2011}. One should also expect drifts from the lower
latitude spot band to the higher one, with opposite direction of the
trend in the phase-time plots. These are indeed seen, but with less
pronounced phase coherence.

In \citet{alekseev2005} it was also proposed, based on photometric spot
modeling, that a latitudinal dynamo wave, the spot activity migrating
from the equator poleward (the solar butterfly reversed), is present
on the object. The analysis of \citet{lehtinen2012} did not reveal
such trends, nor does the CF analysis picture of linear down- or
upward trends support this picture.

The phase behavior, if not due to differential rotation nor
latitudinal dynamo waves, could also be a manifestation of an
azimuthal dynamo wave, predicted to be excited in rapid rotators
\citep[e.g.][]{KR80}, verified from mean-field dynamo models
\citep[e.g.][]{moss1995,kuker1999,mantere2013}, and now also found
from direct numerical simulations \citep{cole2014a}. Such dynamo waves
most often behave as if detached from the overall rotation of the
object, moving with a different speed than the stellar surface. Their
rotation is rigid even in a differentially rotating object. Therefore,
a systematic linear phase drift could most directly be linked to the
presence of azimuthal dynamo waves. Dynamo theory, on the other hand,
serves no direct explanation as to why the linear trends are broken
and reversed, which is clearly the case for LQ Hya. 

\section{Conclusions}\label{conc}

In this work we have presented analysis of LQ Hya photometry for 1982-2014. 
Several different statistical methods were first used to nail down a suitable carrier frequency for our main analysis tool, the CF method.
From this preliminary analysis we learned several interesting aspects:

Firstly, there is a certain cut-off in the spectrum at the high frequency end. 
This can be interpreted as a limiting value for the spot cycle length at the low latitudes or near the equator of the star. 
Second interesting feature appearing as the result of the same analysis is the bimodal shape of the spectrum. 
The explanation for this can be searched from different causes e.g. latitudinal distribution of the spots. From DI maps for LQ Hya 
it has become evident that spot regions tend to lie either on high or low latitudes while there seems to be spotless area 
on mid-latitude range \citep{cole2014b,donati2003}. Other possible explanations might be either 
radial differential rotation manifesting itself through
different anchoring depths for spots 
or hemispherical asymmetry.

In previous studies the focus has been on searching for active longitudes on the star \citep{jetsu1993,berdyugina2002,kovari2004,lehtinen2012}.
Here we took a different approach by estimating the mean rotation period of the spot structures on the star using the phase dispersion statistic $D^2(P)$. 
This period is a close analogue to the Carrington rotation period of the Sun. In subsequent CF analysis we used the obtained value
as a carrier period and produced the corresponding phase plots. We noticed shorter and longer, nearly linear, trends with different slopes reflecting
the "duty times" of certain periods during these time frames. 
Especially pronounced are two epochs (1991--1993; 2005--2008) with a downward trend.
The overall picture is inconsistent with the antisolar butterfly diagram postulated by \citet{alekseev2005}.
The possible sources of these trends include disrupted azimuthal dynamo waves and solar-like latitudinal differential rotation.

We also tried a multiperiodic model to describe the photometry of LQ Hya. This however led to low $R^2$ values and the regular structure 
of the simple model was lost in the phase diagram of the CF. Therefore we concluded this model to be barely suitable.
Interestingly enough, this analysis gave a phase modulation period of roughly 6.94 years, a value also derived from the mean brightness
variation of the star.

From the global CF we calculated the phases of the minima and compared them with the results obtained by \citet{lehtinen2012} using the CPS method.
These two models appeared to be in good agreement. Comparison with the results by \citet{cole2014b} using DI technique was challenging due to the non-overlapping
of the corresponding observing seasons. However, around late 1999 and late 2000 rough agreement between the epochs of the 
photometric minima and the DI spots can be seen.

Qualitative analysis of the local CF was done for 27 observing seasons. We detected 4 flip-flop type events from the phase plots, 3 of these 
matching with the epochs of flip-flops obtained from the global CF. The timing of the events appears to be random which excludes the possibility 
of the 5.2 year cycle reported by \citet{berdyugina2002}. Comparison of the phase plots with the ones reported by \citet{lehtinen2012} 
showed a good agreement -- majority of features can be detected from both analysis. 

\newcommand{\etal}{et al.}

\begin{acknowledgements}
  This work has been supported by the Academy of Finland Centre of
  Excellence ReSoLVE (NO, MM, JP). The work of TH was financed through the
  project Active Suns by the University of Helsinki.
  G.W.H. acknowledges 
  long-term support from Tennessee State University and the State of Tennessee 
  through its Centers of Excellence program.
\end{acknowledgements}

\bibliographystyle{aa}
\bibliography{lqhya}

\begin{thebibliography}{47}
\expandafter\ifx\csname natexlab\endcsname\relax\def\natexlab#1{#1}\fi

\bibitem[{{Alekseev}(2005)}]{alekseev2005}
{Alekseev}, I.~Y. 2005, Astrophysics, 48, 20

\bibitem[{{Barnes} {et~al.}(2005){Barnes}, {Collier Cameron}, {Donati},
  {James}, {Marsden}, \& {Petit}}]{barnes2005}
{Barnes}, J.~R., {Collier Cameron}, A., {Donati}, J.-F., {et~al.} 2005, \mnras,
  357, L1

\bibitem[{{Barning}(1963)}]{barning1963}
{Barning}, F.~J.~M. 1963, \bain, 17, 22

\bibitem[{{Berdyugina} {et~al.}(2002){Berdyugina}, {Pelt}, \&
  {Tuominen}}]{berdyugina2002}
{Berdyugina}, S.~V., {Pelt}, J., \& {Tuominen}, I. 2002, \aap, 394, 505

\bibitem[{{Cole} {et~al.}(2014{\natexlab{a}}){Cole}, {Hackman}, {K\"apyl\"a},
  {Ilyin}, {Kochukhov}, \& {Piskunov}}]{cole2014b}
{Cole}, E., {Hackman}, T., {K\"apyl\"a}, M.~J., {et~al.} 2014{\natexlab{a}},
  \aap, submitted

\bibitem[{{Cole} {et~al.}(2014{\natexlab{b}}){Cole}, {K{\"a}pyl{\"a}},
  {Mantere}, \& {Brandenburg}}]{cole2014a}
{Cole}, E., {K{\"a}pyl{\"a}}, P.~J., {Mantere}, M.~J., \& {Brandenburg}, A.
  2014{\natexlab{b}}, \apjl, 780, L22

\bibitem[{{Covino} {et~al.}(2001){Covino}, {Panzera}, {Tagliaferri}, \&
  {Pallavicini}}]{covino2001}
{Covino}, S., {Panzera}, M.~R., {Tagliaferri}, G., \& {Pallavicini}, R. 2001,
  \aap, 371, 973

\bibitem[{{Cutispoto}(1991)}]{cutispoto1991}
{Cutispoto}, G. 1991, \aaps, 89, 435

\bibitem[{{Cutispoto}(1998)}]{cutispoto1998}
{Cutispoto}, G. 1998, \aaps, 131, 321

\bibitem[{{Donati}(1999)}]{donati1999}
{Donati}, J.-F. 1999, \mnras, 302, 457

\bibitem[{{Donati} {et~al.}(2003{\natexlab{a}}){Donati}, {Collier Cameron}, \&
  {Petit}}]{donati2003b}
{Donati}, J.-F., {Collier Cameron}, A., \& {Petit}, P. 2003{\natexlab{a}},
  \mnras, 345, 1187

\bibitem[{{Donati} {et~al.}(2003{\natexlab{b}}){Donati}, {Collier Cameron},
  {Semel}, {Hussain}, {Petit}, {Carter}, {Marsden}, {Mengel}, {L{\'o}pez
  Ariste}, {Jeffers}, \& {Rees}}]{donati2003}
{Donati}, J.-F., {Collier Cameron}, A., {Semel}, M., {et~al.}
  2003{\natexlab{b}}, \mnras, 345, 1145

\bibitem[{{Eggen}(1984)}]{eggen1984}
{Eggen}, O.~J. 1984, \aj, 89, 1358

\bibitem[{{Fekel} {et~al.}(1986){Fekel}, {Bopp}, {Africano}, {Goodrich},
  {Palmer}, {Quingley}, \& {Simon}}]{fekel1986}
{Fekel}, F.~C., {Bopp}, B.~W., {Africano}, J.~L., {et~al.} 1986, \aj, 92, 1150

\bibitem[{{Gunn} {et~al.}(1998){Gunn}, {Mitrou}, \& {Doyle}}]{gunn1998}
{Gunn}, A.~G., {Mitrou}, C.~K., \& {Doyle}, J.~G. 1998, \mnras, 296, 150

\bibitem[{{Hackman} {et~al.}(2013){Hackman}, {Pelt}, {Mantere}, {Jetsu},
  {Korhonen}, {Granzer}, {Kajatkari}, {Lehtinen}, \&
  {Strassmeier}}]{hackman2013}
{Hackman}, T., {Pelt}, J., {Mantere}, M.~J., {et~al.} 2013, \aap, 553, A40

\bibitem[{{Jetsu}(1993)}]{jetsu1993}
{Jetsu}, L. 1993, \aap, 276, 345

\bibitem[{{Kim} \& {Demarque}(1996)}]{kim1996}
{Kim}, Y.-C. \& {Demarque}, P. 1996, \apj, 457, 340

\bibitem[{{Kitchatinov} \& {Olemskoy}(2011)}]{kitchatinov2011}
{Kitchatinov}, L.~L. \& {Olemskoy}, S.~V. 2011, \mnras, 411, 1059

\bibitem[{{Kov{\'a}ri} {et~al.}(2004){Kov{\'a}ri}, {Strassmeier}, {Granzer},
  {Weber}, {Ol{\'a}h}, \& {Rice}}]{kovari2004}
{Kov{\'a}ri}, Z., {Strassmeier}, K.~G., {Granzer}, T., {et~al.} 2004, \aap,
  417, 1047

\bibitem[{{Krause} \& {Raedler}(1980)}]{KR80}
{Krause}, F. \& {Raedler}, K.~H. 1980, {Mean-field magnetohydrodynamics and
  dynamo theory}

\bibitem[{{K{\"u}ker} \& {R{\"u}diger}(1999)}]{kuker1999}
{K{\"u}ker}, M. \& {R{\"u}diger}, G. 1999, \aap, 346, 922

\bibitem[{{Lehtinen} {et~al.}(2011){Lehtinen}, {Jetsu}, {Hackman}, {Kajatkari},
  \& {Henry}}]{lehtinen2011}
{Lehtinen}, J., {Jetsu}, L., {Hackman}, T., {Kajatkari}, P., \& {Henry}, G.~W.
  2011, \aap, 527, A136

\bibitem[{{Lehtinen} {et~al.}(2012){Lehtinen}, {Jetsu}, {Hackman}, {Kajatkari},
  \& {Henry}}]{lehtinen2012}
{Lehtinen}, J., {Jetsu}, L., {Hackman}, T., {Kajatkari}, P., \& {Henry}, G.~W.
  2012, \aap, 542, A38

\bibitem[{{Lindborg} {et~al.}(2011){Lindborg}, {Korpi}, {Hackman}, {Tuominen},
  {Ilyin}, \& {Piskunov}}]{lindborg2011}
{Lindborg}, M., {Korpi}, M.~J., {Hackman}, T., {et~al.} 2011, \aap, 526, A44

\bibitem[{{Lindborg} {et~al.}(2013){Lindborg}, {Mantere}, {Olspert}, {Pelt},
  {Hackman}, {Henry}, {Jetsu}, \& {Strassmeier}}]{lindborg2013}
{Lindborg}, M., {Mantere}, M.~J., {Olspert}, N., {et~al.} 2013, \aap, 559, A97

\bibitem[{{Lomb}(1976)}]{lomb1976}
{Lomb}, N.~R. 1976, \apss, 39, 447

\bibitem[{{Mantere} {et~al.}(2013){Mantere}, {K{\"a}pyl{\"a}}, \&
  {Pelt}}]{mantere2013}
{Mantere}, M.~J., {K{\"a}pyl{\"a}}, P.~J., \& {Pelt}, J. 2013, in IAU
  Symposium, Vol. 294, IAU Symposium, ed. A.~G. {Kosovichev}, E.~{de Gouveia
  Dal Pino}, \& Y.~{Yan}, 175--186

\bibitem[{{McIvor} {et~al.}(2004){McIvor}, {Jardine}, {Collier Cameron},
  {Wood}, \& {Donati}}]{mcivor2004}
{McIvor}, T., {Jardine}, M., {Collier Cameron}, A., {Wood}, K., \& {Donati},
  J.-F. 2004, \mnras, 355, 1066

\bibitem[{{Messina} \& {Guinan}(2003)}]{messina2003}
{Messina}, S. \& {Guinan}, E.~F. 2003, \aap, 409, 1017

\bibitem[{{Moss} {et~al.}(1995){Moss}, {Barker}, {Brandenburg}, \&
  {Tuominen}}]{moss1995}
{Moss}, D., {Barker}, D.~M., {Brandenburg}, A., \& {Tuominen}, I. 1995, \aap,
  294, 155

\bibitem[{{Ol{\'a}h} {et~al.}(2009){Ol{\'a}h}, {Koll{\'a}th}, {Granzer},
  {Strassmeier}, {Lanza}, {J{\"a}rvinen}, {Korhonen}, {Baliunas}, {Soon},
  {Messina}, \& {Cutispoto}}]{olah2009}
{Ol{\'a}h}, K., {Koll{\'a}th}, Z., {Granzer}, T., {et~al.} 2009, \aap, 501, 703

\bibitem[{{Ol{\'a}h} {et~al.}(2000){Ol{\'a}h}, {Koll{\'a}th}, \&
  {Strassmeier}}]{olah2000}
{Ol{\'a}h}, K., {Koll{\'a}th}, Z., \& {Strassmeier}, K.~G. 2000, \aap, 356, 643

\bibitem[{{Ossendrijver}(1997)}]{Ossendrijver1997}
{Ossendrijver}, A.~J.~H. 1997, \aap, 323, 151

\bibitem[{{Pelt}(1983)}]{pelt1983}
{Pelt}, J. 1983, in ESA Special Publication, Vol. 201, Statistical Methods in
  Astronomy, ed. E.~J. {Rolfe}, 37--42

\bibitem[{{Pelt} {et~al.}(2011){Pelt}, {Olspert}, {Mantere}, \&
  {Tuominen}}]{pelt2011}
{Pelt}, J., {Olspert}, N., {Mantere}, M.~J., \& {Tuominen}, I. 2011, \aap, 535,
  A23

\bibitem[{{Reinhold} \& {Reiners}(2013)}]{reinhold2013}
{Reinhold}, T. \& {Reiners}, A. 2013, \aap, 557, A11

\bibitem[{{Rice} \& {Strassmeier}(1998)}]{rice1998}
{Rice}, J.~B. \& {Strassmeier}, K.~G. 1998, \aap, 336, 972

\bibitem[{{Saar} \& {Brandenburg}(1999)}]{Saar1999}
{Saar}, S.~H. \& {Brandenburg}, A. 1999, \apj, 524, 295

\bibitem[{{Saar} {et~al.}(1994){Saar}, {Piskunov}, \& {Tuominen}}]{saar1994}
{Saar}, S.~H., {Piskunov}, N.~E., \& {Tuominen}, I. 1994, in Astronomical
  Society of the Pacific Conference Series, Vol.~64, Cool Stars, Stellar
  Systems, and the Sun, ed. J.-P. {Caillault}, 661--663

\bibitem[{{Scargle}(1982)}]{scargle1982}
{Scargle}, J.~D. 1982, \apj, 263, 835

\bibitem[{{Stellingwerf}(1978)}]{stellingwerf}
{Stellingwerf}, R.~F. 1978, \apj, 224, 953

\bibitem[{{Strassmeier} {et~al.}(1997){Strassmeier}, {Bartus}, {Cutispoto}, \&
  {Rodono}}]{strassmeier1997}
{Strassmeier}, K.~G., {Bartus}, J., {Cutispoto}, G., \& {Rodono}, M. 1997,
  \aaps, 125, 11

\bibitem[{{Strassmeier} {et~al.}(1993){Strassmeier}, {Rice}, {Wehlau}, {Hill},
  \& {Matthews}}]{strassmeier1993}
{Strassmeier}, K.~G., {Rice}, J.~B., {Wehlau}, W.~H., {Hill}, G.~M., \&
  {Matthews}, J.~M. 1993, \aap, 268, 671

\bibitem[{{Tetzlaff} {et~al.}(2011){Tetzlaff}, {Neuh{\"a}user}, \&
  {Hohle}}]{Tetzlaff2011}
{Tetzlaff}, N., {Neuh{\"a}user}, R., \& {Hohle}, M.~M. 2011, \mnras, 410, 190

\bibitem[{{White} {et~al.}(2007){White}, {Gabor}, \& {Hillenbrand}}]{white2007}
{White}, R.~J., {Gabor}, J.~M., \& {Hillenbrand}, L.~A. 2007, \aj, 133, 2524

\bibitem[{{You}(2007)}]{you2007}
{You}, J. 2007, \aap, 475, 309

\end{thebibliography}

\end{document}